\documentclass{aastex62}

\newcommand\target{MAXI~J1752$-$457~}

\usepackage{multirow}
\usepackage{CJKutf8, here}
\usepackage{amsmath}

\received{}
\revised{}
\accepted{}
\submitjournal{ApJL}

\shorttitle{Thermonuclear superburst of MAXI~J1752$-$457}
\shortauthors{Aoyama \& Enoto et al.}

\begin{document}

\title{Thermonuclear superburst of MAXI~J1752$-$457 observed with NinjaSat and MAXI}

\correspondingauthor{Amira Aoyama, Teruaki Enoto}
\email{amira.aoyama@riken.jp, enoto.teruaki.2w@kyoto-u.ac.jp}

\newcommand{\TUS}{Department of Physics, Tokyo University of Science, 1-3 Kagurazaka, Shinjuku, Tokyo 162-8601, Japan}
\newcommand{\RIKENCPR}{RIKEN Cluster for Pioneering Research (CPR), 2-1 Hirosawa, Wako, Saitama 351-0198, Japan}
\newcommand{\RIKENPRI}{RIKEN Pioneering Research Institute (PRI), 2-1 Hirosawa, Wako, Saitama 351-0198, Japan}
\newcommand{\RIKENRAP}{RIKEN Center for Advanced Photonics (RAP), 2-1 Hirosawa, Wako, Saitama 351-0198, Japan}
\newcommand{\RIKENNishina}{RIKEN Nishina Center, 2-1 Hirosawa, Wako, Saitama 351-0198, Japan}
\newcommand{\KyotoU}{Department of Physics, Kyoto University, Kitashirakawa Oiwake, Sakyo, Kyoto 606-8502, Japan}
\newcommand{\ChibaU}{International Center for Hadron Astrophysics, Chiba University, 1-33 Yayoi, Inage, Chiba 263-8522, Japan}
\newcommand{\iTHEMS}{RIKEN Center for Interdisciplinary Theoretical \& Mathematical Sciences (iTHEMS), RIKEN 2-1 Hirosawa, Wako, Saitama 351-0198, Japan}
\newcommand{\NCUE}{Department of Physics, National Changhua University of Education (NCUE), Changhua 50007, Taiwan}
\newcommand{\CNS}{Center for Nuclear Study (CNS), The University of Tokyo, 7-3-1 Hongo, Bunkyo, Tokyo 113-0033, Japan}
\newcommand{\NAOJ}{National Astronomical Observatory of Japan (NAOJ), 2-21-1 Osawa, Mitaka, Tokyo 181-8588 , Japan}
\newcommand{\AGU}{Department of Physical Science, Aoyama Gakuin University, 5-10-1 Fuchinobe, Chuo-ku, Sagamihara, Kanagawa 252-5258, Japan}
\newcommand{\NihonU}{School of Dentistry at Matsudo, Nihon University, 2-870-1, Sakaecho-nishi, Matsudo, Chiba 271-8587, Japan}

\author[0009-0008-3926-363X]{Amira Aoyama} 
\affiliation{\TUS}
\affiliation{\RIKENPRI}

\author[0000-0003-1244-3100]{Teruaki Enoto}
\affiliation{\KyotoU}
\affiliation{\RIKENRAP}

\author{Takuya Takahashi}
\affiliation{\TUS}
\affiliation{\RIKENNishina}

\author{Sota Watanabe}
\affiliation{\TUS}
\affiliation{\RIKENNishina}

\author[0009-0008-5133-9131]{Tomoshi Takeda}
\affiliation{\TUS}
\affiliation{\RIKENPRI}

\author[0000-0002-0207-9010]{Wataru Iwakiri}
\affiliation{\ChibaU}

\author{Kaede Yamasaki}
\affiliation{\TUS}
\affiliation{\RIKENNishina}

\author{Satoko Iwata}
\affiliation{\TUS}
\affiliation{\RIKENNishina}

\author{Naoyuki Ota}
\affiliation{\TUS}
\affiliation{\RIKENNishina}

\author{Arata Jujo}
\affiliation{\TUS}
\affiliation{\RIKENNishina}

\author[0000-0002-8801-6263]{Toru Tamagawa}
\affiliation{\RIKENPRI}
\affiliation{\RIKENNishina}
\affiliation{\TUS}

\author{Tatehiro Mihara}
\affiliation{\RIKENPRI}

\author[0000-0001-8551-2002]{Chin-Ping Hu}
\affiliation{\NCUE}

\author[0000-0001-8726-5762]{Akira Dohi}
\affiliation{\RIKENPRI}
\affiliation{\iTHEMS}

\author[0000-0002-0842-7856]{Nobuya Nishimura}
\affiliation{\CNS}
\affiliation{\RIKENPRI}
\affiliation{\NAOJ}

\author{Motoko Serino}
\affiliation{\AGU}

\author{Motoki Nakajima}
\affiliation{\NihonU}

\author{Takao Kitaguchi}
\affiliation{\RIKENPRI}

\author{Yo Kato}
\affiliation{\RIKENPRI}

\author{Nobuyuki Kawai}
\affiliation{\RIKENPRI}

\collaboration{(NinjaSat collaboration)}

\begin{abstract}
An uncatalogued bright X-ray transient was detected with MAXI on November 9, 2024, named MAXI~J1752$-$457. 
The NinjaSat X-ray observatory promptly observed the source from November 10 to 18 while the small angular separation from the Sun hampered follow-up campaigns by other X-ray observatories.
The MAXI and NinjaSat light curves in the 2--10 keV band showed first and second decaying components at the early and late phases, approximated by exponential functions with e-folding constants of 1.2 $\pm$ 0.2 and 14.9 $\pm$ 0.9~hours (1$\sigma$ errors), respectively. 
A single blackbody model reproduces the X-ray spectrum with a softening trend of its temperature decreasing from 1.8 $\pm$ 0.1 keV to 0.59 $\pm$ 0.06 keV. Assuming the unknown source distance at 8~kpc, at which the initial X-ray luminosity roughly corresponds to the Eddington limit, the shrinking blackbody radius was estimated at 5--11~km. 
This X-ray brightening is interpreted as a superburst in a Galactic low-mass X-ray binary, which is powered by thermonuclear burning triggered presumably by carbon ignition close to the outer crust of the neutron star.
%in the neutron-star outer crust. 
The transition between two decaying components occurred at 5.5--7.7~hours, corresponding to the thermal time scale of the burning layer. The ignition column density is estimated to be (2.8--5.1)$\times 10^{12}$~g~cm$^{-2}$.
\end{abstract}


%------------------------
\section{Introduction}
\label{sec:Introduction}
%------------------------

Accreting neutron stars (NSs)  in low-mass X-ray binaries often become bright as an explosive transient triggered by thermonuclear runaway, called Type-I X-ray bursts, on a normally minute timescale. However, they sometimes show a long duration of several hours, three orders of magnitude longer than normal X-ray bursts, called superbursts. It is widely believed that superbursts are caused by the ignition of carbon close to the outer crust \citep{2021ASSL..461..209G} with the high stellar-crust column density of $y\sim10^{12}~{\rm g~cm^{-1}}$ compared with the low density of typical X-ray bursts ($y\sim10^8~{\rm g~cm^{-2}}$) of hydrogen burning\footnote{Another type of long X-ray burst, intermediate long burst, is triggered by the burning of helium layers inside accreting NSs.}. X-ray light curves of superbursts can provide information on the thermal and compositional structure of accreting NSs. Until now, 16 superburst sources have been reported in the recent catalog \citep{2023MNRAS.521.3608A}.
%In this paper, we report the discovery of the 18th superburster, \target.}

\begin{figure}[ht!]
\centering
\includegraphics[width=0.9\textwidth]{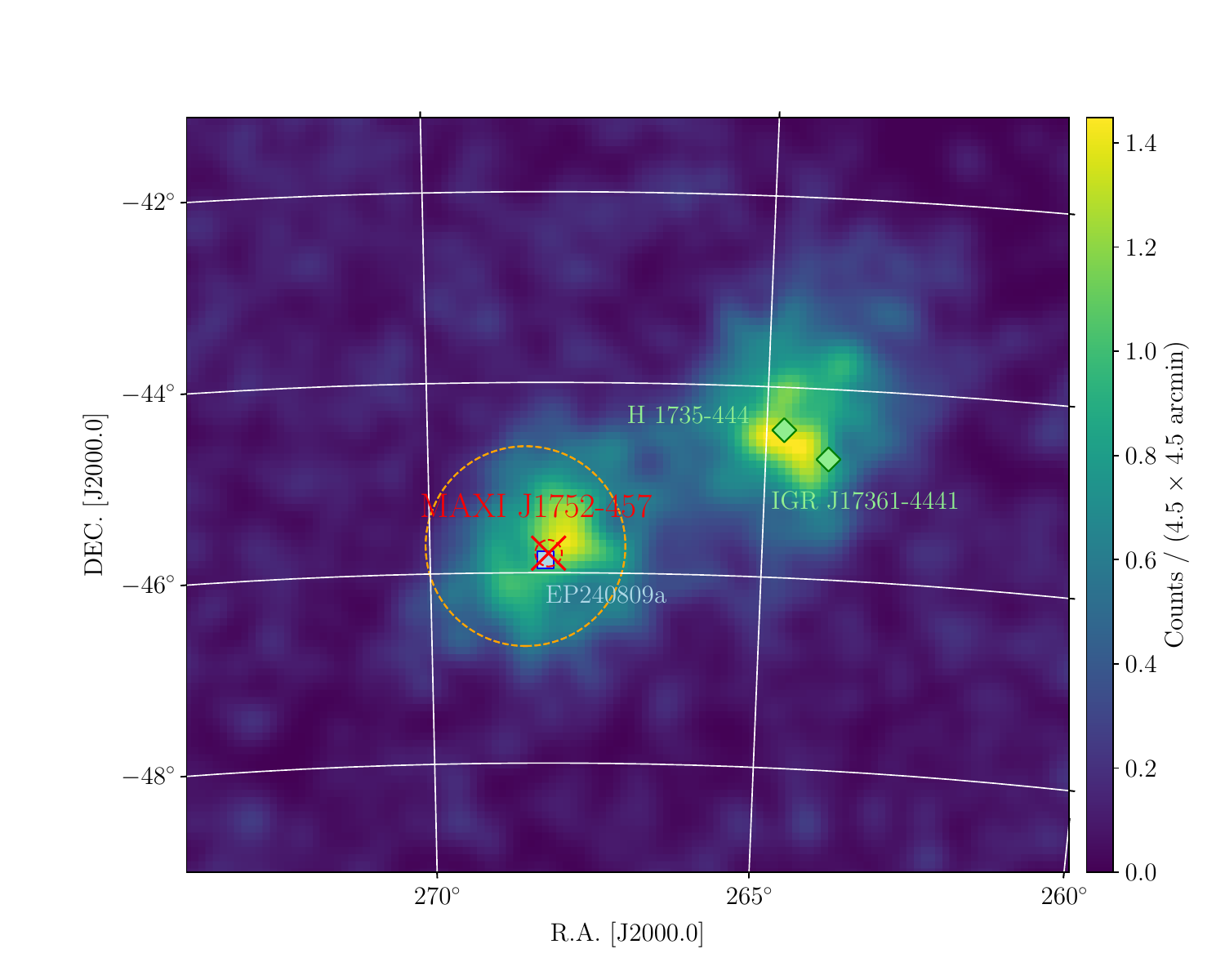}
\caption{The MAXI/GSC 2.0--10.0 keV image obtained between MJD 60623.5 and 60624.0 after being smoothed with a Gaussian filter of its Gaussian kernel of 2~pixel (9.0~arcmin). The color map is shown in a unit of integrated X-ray counts at each 4.5 $\times$ 4.5~arcmin pixel. The positions of \target reported by MAXI and EP240809a by Swift are shown as the red cross and blue square markers, respectively~\citep{2024ATel16898....1S,2024ATel16765....1L}. The MAXI 1$\sigma$ error region is shown as the dashed red circle, whereas the reported EP240809a (Swift) error circle is invisibly small in this plot. 
The NinjaSat GMC boresight direction and the 2.1$^{\circ}$ FoV are shown in the orange cross and orange dashed circle, respectively. 
The nearby X-ray sources H~1735$-$444 and IGR~J17361$-$4441 (green diamonds) are outside of the NinjaSat FoV.
\label{fig:position}
}
\end{figure}

On November 9, 2024, at 18:23 UT, the Monitor of All-sky X-ray Image (MAXI; \citealt{2009PASJ...61..999M}) detected an uncatalogued X-ray transient with the Gas Slit Camera (GSC) at the J2000 source position of (R.A., Dec.)=(268.213$^{\circ}$, $-45.795^{\circ}$), and it was named MAXI~J1752$-$457~\citep{2024ATel16898....1S}. 
The X-ray flux declined exponentially with a time constant of $0.10 \pm 0.03$ days and $0.08 \pm 0.01$ days in the 2--4~keV and 4--10~keV band, respectively, indicating spectral softening during the decay~\citep{2024ATel16902....1N}.
The spectra were described with an absorbed blackbody model, whose temperature decreased from $1.8^{+0.2}_{-0.1}$~keV to $1.2^{+0.2}_{-0.2}$~keV in 6~hours.
On November 12--13, 2024, a NuSTAR follow-up observation determined the \target position at (R.A., Dec.)=(268.269$^{\circ}$,$-45.866^{\circ}$), which is only 9~arcsec apart from the X-ray source EP240809a, previously identified by the X-Ray Telescope (XRT) onboard Swift (\citealt{2024ATel16765....1L,2024ATel16767....1Z, 2024ATel16910....1P}). 
Figure~\ref{fig:position} shows the MAXI GSC 2--10~keV image around \target and EP240809a. Considering the localization uncertainties of NuSTAR (18~arcsec point spread function, FWHM) and Swift (3.7~arcsec; 
\citealt{2024ATel16765....1L}), \target is likely the same source as EP240809a.

Between the first two MAXI scans on November 9, the hour-scale decline in X-ray flux of \target suggested the need for prompt and continuous monitoring by pointing X-ray observatories. 
However, the Sun angle constraint hindered the Target-of-Opportunity (ToO) observations of most X-ray observatories, including Swift and NICER, since \target was only 45$^{\circ}$ from the Sun. %during this season.
The NinjaSat X-ray observatory is a 6U-size CubeSat launched on November 11, 2023~\citep{Enoto2020,Tamagawa2024}.
Since NinjaSat has a more relaxed Sun angle constraint compared to other X-ray satellites, we promptly started a ToO observation by NinjaSat on November 10 at 12:53:55 UT (MJD~60624.54). 
It was approximately only 2.5~hours after the initial ATel post and 1.69~days after the detection by MAXI~\citep{2024ATel16903....1T}. 
This campaign continued until November 18, when the Sun angle constraint made further NinjaSat observations challenging.

% In this Letter, we report NinjaSat and MAXI observations of \target and its interpretation.
In this Letter, we report NinjaSat and MAXI observations of \target and its plausible interpretation of the superburst. Quoted errors represent the 1$\sigma$ confidence interval.

%------------------------
\section{Observation and Data Reduction}
%------------------------
As of November 11, 2024, NinjaSat was orbiting at an approximate altitude of 480~km in the Sun-synchronous orbit.
10\%  of the time was spent on the pointing observation of MAXI~J1752$-$457, excluding the time while communicating with the ground stations, and in the high particle background regions such as the south Atlantic anomaly and high-latitude auroral zones.
NinjaSat is equipped with two non-imaging Xe-based Gas Multiplier Counters (GMCs) covering the 2--50~keV band with a total 32~cm$^2$ effective area  (two GMCs) at 6~keV with an energy resolution of 19.8\%, FWHM~\citep{Enoto2020,Tamagawa2024}.
We only used GMC1 of the two GMCs for the present observations. 
The scientifically available exposure (Good Time Intervals, GTIs) was 71.2~ks from November 10 to 18 for MAXI~J1752$-$457. 
Since we can observe multiple targets in a single day, we additionally observed the Crab Nebula from November 9 to 15, and the blank-sky position of (R.A., Dec.)=(138.00$^{\circ}$, 15.00$^{\circ}$), also known as BKGD\_RXTE3 in the NICER archive~\citep{2022AJ....163..130R}, from November 15 to 29, to verify the data quality and background studies.
The GTI exposures of the Crab Nebula and the blank-sky region were 68.0 and 114.3~ks, respectively. 

The X-ray collimator of the GMC limits the field-of-view (FoV) of NinjaSat to 2.1$^{\circ}$ at the full-width at half-maximum (FWHM). The GMC boresight position and FoV are shown in Figure \ref{fig:position} with nearby bright X-ray sources. Our detailed analyses of the satellite attitude data determined the mean target direction of the GMC1 boresight axis at (R.A., Dec.)=($268.56^{\circ}$, $-45.72^{\circ}$), slightly shifted by $0.25^{\circ}$ from the \target position. 
This offset reduces the X-ray intensity by 16\% from the boresight value due to the open aperture ratio of the X-ray collimator. This 16\% reduction is included in the following spectral and light curve analyses. The time fluctuation of the boresight directions at the R.A. and Dec. directions is  0.12$^{\circ}$ and $0.17^{\circ}$ in standard deviation, respectively, which are small enough relative to the collimator response. Two nearby bright X-ray sources in the MAXI image, H~1735$-$444 (20~mCrab) and IGR~J17361$-$4441 (1.6~mCrab), are 2.98$^{\circ}$ and 3.31$^{\circ}$ apart from MAXI~J1752$-$457, and contamination outside the NinjaSat FoV is negligible. 
We also checked the Swift XRT 0.3--10~keV image (TargetIDs 00016764) obtained from August 8 to October 24, 15 days before \target appeared.
There was no X-ray point source brighter than 0.37~mCrab within the FoV of NinjaSat around MAXI~J1752$-$457.
%within an 8.4~arcmin radius around MAXI~J1752$-$457.

Following the method of \citet{2024arXiv241110992T}, we extracted GMC cleaned events of MAXI~J1752$-$457. The count rate in the 2--10 keV band was 0.655~counts~s$^{-1}$ on November 10 and decreased to the background level of 0.244~counts~s$^{-1}$ by November 14. The event rates of the Crab Nebula and the blank sky were $11.95\pm0.02$~counts~s$^{-1}$ and $0.244\pm0.002$~counts~s$^{-1}$, respectively. %$0.2439\pm0.002$

We investigated the background characteristics using the blank sky observations of BKGD\_RXTE3, which is mainly composed of non-X-ray background and cosmic X-ray background. 
After applying the same GTI filtering and event cut as MAXI~J1752$-$457, we extracted the 1-day binned 2--10~keV count rate of the background. 
The mean rate and standard deviation of the cleaned data are 0.240~counts~s$^{-1}$ and 0.005~counts~s$^{-1}$, respectively.
Thus, the average background level is 24~mCrab X-ray intensity, and its 1$\sigma$ fluctuation is 0.5~mCrab. %(1$\sigma$) level. 
Figure~\ref{fig:lc} shows the light curve of this campaign.
In this figure, we subtracted the averaged background rate from the \target rate and binned the data at a 1-day or longer time bin in the late phase. 
We evaluated the detection sensitivity of NinjaSat to be the 3$\sigma$ fluctuation (about 0.015~counts~s$^{-1}$) of the background rate, which is 1.5~mCrab or $3.6\times 10^{-11}$~ergs~s$^{-1}$~cm$^{-2}$ for 8~ks in the present background subtraction procedure. 

In order to perform X-ray spectral fitting of NinjaSat, we generated XSPEC response files composed of the redistribution matrix file (rmf) of the gas detector and the auxiliary response file (arf) of the X-ray collimator and the quantum efficiency of the detector. 
The former rmf is constructed based on the Geant4 particle simulation~\citep{Agostinelli2003} that incorporates the detector geometry and physical interactions in the gas cell. We also incorporate other effects, such as electron transport and amplification in the gas cell, the circuit response of the front-end card and data acquisition board, and the event screening based on information of X-ray signals. 
We verified the arf through the Crab Nebula raster scan observations and took into account the off-axis correction of the boresight direction. 

We checked the validity of the NinjaSat response files by spectral fitting of the 2--20~keV Crab Nebula obtained from November 9 to 15, 2024 (MJD 60623--60629).
As a reference, we used the NuSTAR absolute flux measurement of the Crab Nebula, which reported the absorbed power-law, ``{\tt tbabs} $\times$ {\tt powerlaw}" in XSPEC, where {\tt tbabs} is the ISM absorption model \citep{2000ApJ...542..914W}, with parameters of the column density of $N_{\textrm{H}}=2.2\times 10^{21}$~cm$^{-2}$, photon index of $\Gamma=2.103$, its normalization of $K=9.69$~ph~s$^{-1}$~cm$^{-2}$~keV$^{-1}$ at 1~keV, and the 2--10~keV X-ray flux of $F_{2-10}=2.09\times 10^{-8}$~ergs~s$^{-1}$~cm$^{-2}$~\citep{Madsen2022}.
Here, we compared our MAXI and NinjaSat spectra with this NuSTAR model.
The normalization of NinjaSat became 17\% smaller than the MAXI and NuSTAR values when fitting them at the fixed column density of $N_{\textrm{H}}=2.2\times 10^{21}$~cm$^{-2}$. 
Thus, we further corrected the effective area of GMC1, reducing it by 17\% in this Letter. 
After including this correction, we derived the best-fit NinjaSat parameters of $\Gamma=2.13 \pm 0.04$, $K=9.8 \pm 0.7$~ph~s$^{-1}$~cm$^{-2}$~keV$^{-1}$ at 1~keV, and $F_{2-10} = (2.08 \pm 0.03)\times 10^{-8}$~ergs~s$^{-1}$~cm$^{-2}$, consistent with NuSTAR.
These parameters are within the previously reported values by other X-ray satellites, e.g., $N_H=(2.9$--$5.4)\times 10^{21}$~cm$^{-2}$, $\Gamma=1.83$--2.25, and $K=7.0$--15.9~ph~s$^{-1}$~cm$^{-2}$~keV$^{-1}$~\citep{Kircsh2005}.
The fitting residuals relative to the best-fit model stay within the 5\% level in the 2--10~keV band.
We also confirmed that MAXI best-fit parameters were consistent with NuSTAR values.

We downloaded the MAXI GSC data from the MAXI on-demand archive ~\citep{Mihara2024} and analyzed it according to the analysis procedure at each time span. 
In Figure \ref{fig:lc}, we divided the time span of the MAXI data into A, B, and C, whereas the NinjaSat data into D to H.

%------------------------
\section{Analysis}
\label{sec:Analysis}
\begin{figure}[ht!]
\centering
\includegraphics[width=0.9\textwidth]{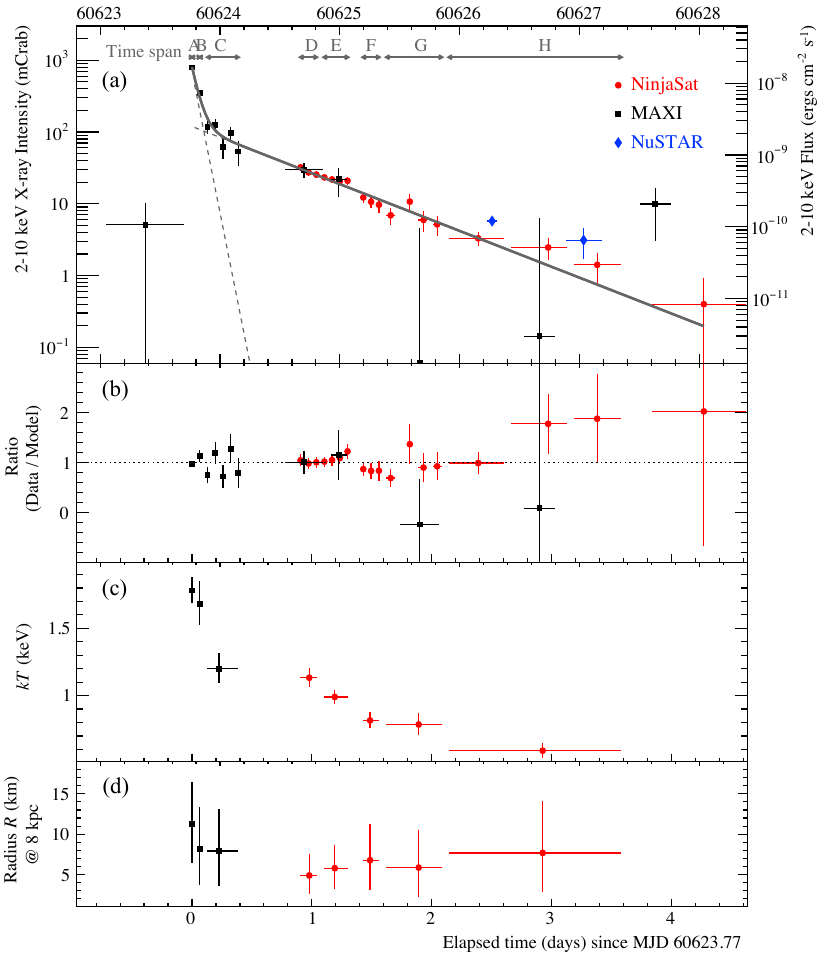}
\caption{
(a) The 2--10 keV light curves of \target observed with NinjaSat/GMC (red circles), MAXI/GSC (black squares), and NuSTAR/XRT (blue diamonds, \citealt{2024ATel16910....1P}) presented in the mCrab unit between MJD 60235 (November 11) and 60632 (November 18). NuSTAR fluxes were taken from~\citet{2024ATel16910....1P}.
The NinjaSat/GMC data are shown with the bin size of 1.5~hours on MJD 60624.6--60626.5, 3.0 hr on MJD 60626.5--60627.4, and 6.0 hr after MJD 60626.5.  
The X-ray intensities with MAXI and NinjaSat are converted to the Crab unit and then to the flux unit (see details in the text).
%unit by using $2.09 \times 10^{-8}$~ergs~cm$^{-2}$~s$^{-1}$ .
%, derived from the Crab spectrum fitting model~\citep{Kirsch2005}. 
%NuSTAR data were converted from X-ray flux to X-ray intensity using this factor. 
Each time span for spectral fittings is indicated in the upper part of the figure.
(b) The ratio of data to the best-fit curve (Eq. \ref{eq:pow}) of panel a (see the main text). 
(c) The black-body temperature. 
(d) The black-body radius of the blackbody model at the assumed distance of 8~kpc (Table \ref{tab:spectral_fitting}). 
\label{fig:lc}
}
\end{figure}

%------------------------
Figure~\ref{fig:lc}a shows the 2--10 keV light curve of \target obtained by MAXI and NinjaSat.
The X-ray count rate is normalized by that of the Crab Nebula to the  ``Crab" unit and then converted to the X-ray flux assuming the conversion factor of 1~Crab unit corresponding to $2.09\times 10^{-8}$~ergs~cm$^{-2}$~s$^{-1}$ in the 2--10~keV band~\citep{Kirsch2005}.
Compared with a short exposure (about 50~s) at each MAXI orbit, NinjaSat provided continuous (about 1000~s) and high-statistics data until November 14 (MJD 60628), when the X-ray intensity declined below the sensitivity limit. 
The X-ray flux at the first MAXI scan on MJD 60623.77 (hereafter $T_0$ November 9, 18:23:02 UTC) is $(2.1 \pm 0.6)\times 10^{-8}$~ergs~s$^{-1}$~cm$^{-2}$ ($982 \pm 50$~mCrab) in the 2--10~keV~\citep{2024ATel16898....1S}. 
Since we do not know the true onset time $t_0$ of this X-ray brightening, which should be within 1.5~hours after the previous MAXI scan at MJD 60623.703, November 9, 16:52:19 UTC (i.e., $60623.70 < t_0 < 60623.77=T_{0}$ in MJD), for simplicity in the following analyses, we assume $t_0=T_0$. 

The X-ray intensity declined by one order of magnitude within 3~hours during the first three MAXI scans (Figure~\ref{fig:lc}a).
The declining slope of the light curve changed at around MJD 60624.0 (November 10), indicating the second slowly decaying component. 
Neither a single power-law model nor a single exponential model can explain the whole curve with chi-square values of $\chi^2=747.3$ (degree of freedom, d.o.f.$=$28) and $\chi^2=172.1$ (d.o.f.$=$28), respectively. 
Thus, we tried two-exponential or two-power-law models. In the former model, the absorbed 2--10~keV X-ray intensity $I(t)$ at time $t$ is written as 
\begin{equation}
I(t)=I_{\textrm{f}}\exp\left(-\frac{t-T_0}{\tau_{\textrm{f}}}\right)+I_{\textrm{s}}\exp\left(-\frac{t-T_0}{\tau_{\textrm{s}}}\right),
\label{eq:exp}
\end{equation}
where $I_{\textrm{f}}$, and $I_{\textrm{s}}$ are the initial X-ray intensities at $T_0$ of the first and second components, of which decay constants (e-folding time) are $\tau_{\textrm{f}}$ and $\tau_{\textrm{s}}$, respectively. 
This model gave an acceptable fit with a chi-square value of 23.6 (d.o.f. = 26) with the best-fit e-folding time of $\tau_{\textrm{f}}=1.2\pm 0.2$~hours and $\tau_{\textrm{s}}=14.9\pm 0.9$~hours. 
The X-ray curve in Figure~\ref{fig:lc} can be successfully modeled by Eq.~\ref{eq:exp} without an additional constant term.
Time constants by~\citet{2024ATel16902....1N} are in-between and consistent since we use two exponential components here.

The other model with two power-law components is written as 
% \begin{equation}
%     I(t)=
%   \begin{cases}
%     I_0 (t-T_0)^{-p_{\textrm{f}}} & \text{$t<t_{\textrm{break}}$,} \\
%     I_0 (t_{\textrm{break}}-T_0)^{-p_{\textrm{f}}}(t-T_0)^{-p_{\textrm{s}}}                  & \text{$t\ge t_{\textrm{break}}$,} 
%   \end{cases}\label{eq:pow}
% \end{equation}
\begin{equation}
    I(t)=
  \begin{cases}
    I_0 (t-T_0)^{-p_{\textrm{f}}} & \text{$t<t_{\textrm{break}}$,} \\
    I_0 (t_{\textrm{break}}-T_0)^{p_{\textrm{s}}-p_{\textrm{f}}}(t-T_0)^{-p_{\textrm{s}}}                  & \text{$t\ge t_{\textrm{break}}$,} 
  \end{cases}\label{eq:pow}
\end{equation}
where $I_0$, $t_{\textrm{break}}$, $p_{\textrm{f}}$, and $p_{\textrm{s}}$ are the initial intensity at $T_0$, the time of the break (change of the slopes), and the first and second decaying slopes, respectively.
This also gave an acceptable fit with a chi-square value of 29.56 (d.o.f=26). The best-fit slopes were $p_{\textrm{f}}=0.88\pm 0.04$ and $p_{\textrm{s}}=3.1 \pm 0.3$, with $t_{\textrm{break}}=1.17 \pm 0.05$~hours. 
The former two-exponential model is slightly better than the latter two-power-law model. 

\begin{figure}[ht!]
\plotone{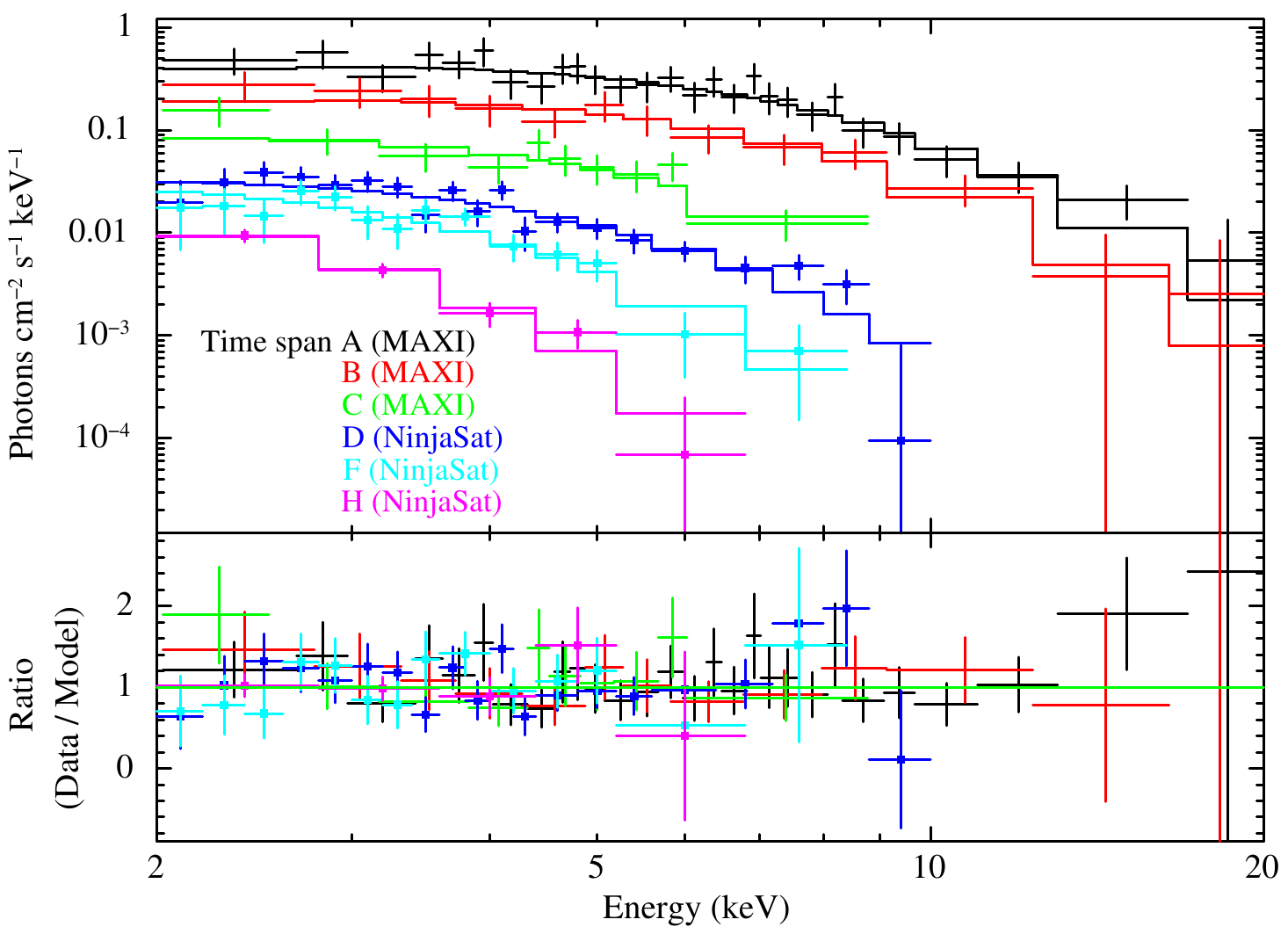}
\caption{{The fitted X-ray spectra of MAXI/GSC and NinjaSat/GMC using the absorbed blackbody model. 
The time spans A, B, and C correspond to MAXI/GSC, with no markers. 
The time spans D, F, and H correspond to NinjaSat/GMC, represented by circular markers. The time spans are shown in Figure \ref{fig:lc}a.
The upper panel displays the unfolded spectrum, while the lower panel shows the ratio of the data to the best-fitting model (see Table \ref{tab:spectral_fitting}).
}
\label{fig:spec}}
\end{figure}

We then performed spectral analyses of MAXI and NinjaSat.
We compared three empirical models: blackbody ({\tt bbody} in the XSPEC terminology), multi-blackbody accretion disk ({\tt diskbb}), and a single power-law ({\tt powerlaw}). 
For all three models, the hydrogen-equivalent column density is fixed to $N_{\rm H}=1\times 10^{21}$~cm$^{-2}$, expected from the HI observations through the \texttt{nh} command in HEASoft (e.g., \citealt{2016A&A...594A.116H}). 
The best-fit parameters are listed in Table~\ref{tab:spectral_fitting} at each time span of the light curve. The blackbody or multi-blackbody disk models reproduce the spectra better than the power-law model. The spectral fit with the single blackbody model is shown in Figure~\ref{fig:spec}, and the time evolution of the model parameters is presented in the panels (c) and (d) of Figure~\ref{fig:lc}.
The bolometric blackbody flux was calculated by using the XSPEC convolutional {\tt cflux} model in the 0.1--100~keV band.

Employing the simple blackbody model, the temperature decreased from $kT=1.8\pm 0.1$ keV at $T_{0}$ to $0.59\pm 0.06$~keV at the late phase. 
When assuming the unknown source distance $d$ at 8~kpc (typical distance to the Galactic center) as the fiducial value, the expected radii of the blackbody component changed from $R=11.2^{+4.8}_{-5.2} (d/8~\textrm{kpc})$~km to $7.7^{+4.8}_{-6.5}(d/8~\textrm{kpc})$~km. 
The disk blackbody model gave 
the inner disk temperature changing from $3.4\pm 0.3$~keV to $0.76 \pm 0.09$~keV, with its disk radius of $2.4^{+1.3}_{-1.5}\cos \theta$~km to $4.2^{+2.8}_{-4.1}\cos \theta$~km, where $\theta$ is the inclination angle of the disk ($\theta=0$ at the face-on geometry).  
The power-law case also showed the softening trend of its photon index from $\Gamma=1.6\pm 0.1$ to $4.2\pm 0.3$. 

% As a timing analysis, we searched for possible periodicity in the cleaned event after applying the barycentric correction (``barycorr" in the HEASoft) at the target UVOT position (R.A., Dec.)=(268.26769$^{\circ}$, -45.86380$^{\circ}$) with the JPL planetary ephemeris DE-430~\citet{2024ATel16765....1L}. 
As a timing analysis, we searched for possible periodicity in the cleaned event after applying the barycentric correction (``barycorr" in the HEASoft) for the position (R.A., Dec.)=(268.26769$^{\circ}$, -45.86380$^{\circ}$), as measured with Swift/UVOT for EP240809a~\citep{2024ATel16765....1L} with the JPL planetary ephemeris DE-430. 
Here, we regrouped the data up to MJD 60628.453 into 21 segments, each corresponding to three GTIs. 
The extracted Fourier power spectrum has no apparent periodical signal with an upper limit of 8\% (2$\sigma$) for all segments in the 0.1--1000 Hz frequency range. 

% -----------------------
\section{Discussion}
% -----------------------
From November 9 to 18, 2024, the X-ray brightening of \target was intensively monitored with MAXI and NinjaSat, from its peak at about 1~Crab X-ray intensity down to the NinjaSat sensitivity limit of 1.5~mCrab. 
Here, we explore possible interpretations of this new transient. 
The observed light curve and spectra differ from classical X-ray brightenings typically seen in the low-mass X-ray binaries (LMXBs) hosting black holes and high-mass X-ray binaries with NSs. %high-mass X-ray binaries with magnetized NSs, or isolated magnetars. 
Classical LMXBs with black holes do not show an X-ray burst. Most of the high-mass X-ray binaries hosting a NS do not show hour-lasting X-ray bursts except for supergiant fast X-ray transients, which exhibit high column density and power-law-like continuum different from the present observations.
%High-mass X-ray binaries with NSs and isolated magnetars do not show long X-ray bursts lasting hours, and the spectra are not a blackbody but a power-law.
Furthermore, the absence of a bright optical companion rules out the possibility of a nearby stellar flare.
When EP240809a (MAXI~J1752$-$457) was first discovered, a tidal disruption event (TDE) was considered as a possible explanation based on a possible blue optical counterpart~\citep{2024ATel16767....1Z}.
However, the observed initial X-ray decay is well approximated by a power-law function with a slope index of $-0.9$, which is inconsistent with the theoretical expectation for a typical TDE with the power-law slope of $-5/3=-1.67$~\citep{Martin1988}.
Moreover, no host galaxy has been identified~\citep{2024ATel16767....1Z}.
Another possibility, a jetted TDE in a distant host galaxy, is also disfavored, as the observed blackbody spectrum differs from the power-law-like shape expected in this scenario, which was also noted in \citet{2024ATel16902....1N}.

A plausible scenario of a thermonuclear burst on an accreting NS in a Galactic LMXB is also suggested by ~\citet{2024ATel16902....1N}. The location near the Galactic bulge is also consistent with the LMXB population~\citep{Fortin:2024siz}.
The blue optical counterpart \citep{2024ATel16765....1L} is also consistent with the LMXB scenario as an irradiated accretion disk.
Adopting the fiducial value $d=8$~kpc, we estimate the 2--10~keV X-ray luminosity to be $1.6^{+0.1}_{-0.1}\times 10^{38}(d/8\, \rm kpc)$~erg~s$^{-1}$ at the peak, reaching the Eddington limit. Alternatatively, we can derive the upper limit on the source distance to be $d<12$~kpc by equating the peak flux with the maximum limit at the Eddington luminosity for a helium photosphere of $3.8\times10^{38}~{\rm erg~s^{-1}}$~\citep{1993SSRv...62..223L}.
% Instead, if the intense photospheric radius expansion burst reached the empirical Eddington limit of $3.8\times10^{38}~{\rm erg~s^{-1}}$~\citep{2003A&A...399..663K}, the observation implies the upper limit on the distance $d<12$ kpc.
In the late phase, the X-ray luminosity declined by three orders of magnitude over approximately three days, reaching the quiescent level of $\sim 10^{35}$~erg~s$^{-1}$.
The derived radius of 5--11~km and temperature of a few keV are consistent with the emission from the NS surface in accreting X-ray binaries. 

Before the present brightening, the Wide-field X-ray Telescope (WXT) onboard EP detected EP240809a on August 9, with an unabsorbed 0.5--4.0~keV X-ray flux of $5.3^{+0.7}_{-0.6}\times 10^{-11}$~ergs~s$^{-1}$~cm$^{-2}$~\citep{2024ATel16765....1L}.
A follow-up observation with Swift/XRT on August 10 measured an unabsorbed 0.3--10~keV X-ray flux of $(1.4\pm 0.1)\times 10^{-11}$~ergs~s$^{-1}$~cm$^{-2}$~\citep{2024ATel16765....1L}.
These fluxes are three orders of magnitude lower than those during our observations.
\citet{2015MNRAS.454.1371W} collected photon indices of known NS-LMXB X-ray spectra fitted by a single power-law model from the literature. They reported a systematic trend; in low luminosity  of $10^{34}$--$10^{36}$~erg~s$^{-1}$, the power-law photon index ranges in $\Gamma=$1.5--3.0 and increases as the X-ray luminosity decreases. The photon indices of MAXI~J1752$-$457, $\Gamma=1.6\pm0.2$ and $1.87\pm 0.14$ reported by EP and Swift, and $\Gamma \sim3$ as observed by NinjaSat in the late phase are consistent with the reported trend for the low-luminosity NS-LMXBs. 

The long e-folding timescales of the first and second decays imply that \target was detected as a long-duration X-ray burst (i.e., superburst). 
The observed blackbody temperature of 0.7--1.8~keV is consistent with typical superburst temperatures~\citep{Serino2016}. 
To investigate the properties of the putative superburst, we utilize the cooling model after an X-ray burst by \citep{Cumming2004}, which gives the analytical formulae of the flux evolution at the NS surface:
\begin{eqnarray}
\hspace*{-1cm}
    F^{\rm NS}(t)&=& \left(2\times10^{24}~{\rm erg~cm^{-2}~s^{-1}}\right) (t-t_0)^{-\alpha_{\rm f}}E_{17}^{7/4}\left[1-\exp\left(-0.63t_{\rm cool}^{4/3}E_{17}^{-5/4}(t-t_0)^{-(\alpha_{\rm s}-\alpha_{\rm f})}\right)\right] \label{eq:cm04} \\ 
    &\simeq&
  \begin{cases}
    F^{\rm NS}_0 (t-t_0)^{-\alpha_{\textrm{f}}} & \text{$t<t_{\textrm{break}}$,} \\
    F^{\rm NS}_0 t_{\textrm{break}}^{\alpha_{\textrm{s}}-\alpha_{\textrm{f}}}(t-t_0)^{-\alpha_{\textrm{s}}}                  & \text{$t\ge t_{\textrm{break}}$,} 
  \end{cases}  \label{eq:cm04_b}
\end{eqnarray}
  % \begin{cases}
  %   F^{\rm NS}_0 (t-t_0)^{-\alpha_{\textrm{f}}} & \text{$t<t_{\textrm{break}}$,} \\
  %   F^{\rm NS}_0 (t_{\textrm{break}}-t_0)^{-\alpha_{\textrm{f}}}(t-t_0)^{-\alpha_{\textrm{s}}}                  & \text{$t\ge t_{\textrm{break}}$,} 
  % \end{cases}  \label{eq:cm04_b}
where $F^{\rm NS}_0$ is the initial bolometric flux at $t=t_0$, and $E_{17}$ denotes the energy released per unit mass of the ignited light element in a unit of $10^{17}~{\rm erg~g^{-1}}$. 
The parameter $t_{\textrm{cool}}$ is the thermal relaxation timescale given as
\begin{eqnarray}
    t_{\rm cool}\approx E_{17}^{1.1}t_{\rm break}\propto y_{12}^{3/4},\label{eq:cm04_2}
\end{eqnarray}
where $y_{12}$ denotes the column density of the ignited layer on the NS surface in a unit of $10^{12}~{\rm g~cm^{-2}}$. 
The slopes of first and second decaying components are predicted to be $\alpha_{\textrm{f}}=0.2$ and $\alpha_{\textrm{s}}=4/3$, respectively, in the standard cooling model~\citep{Cumming2004}.

\begin{figure}[ht!]
\centering
\includegraphics[width=0.9\textwidth]{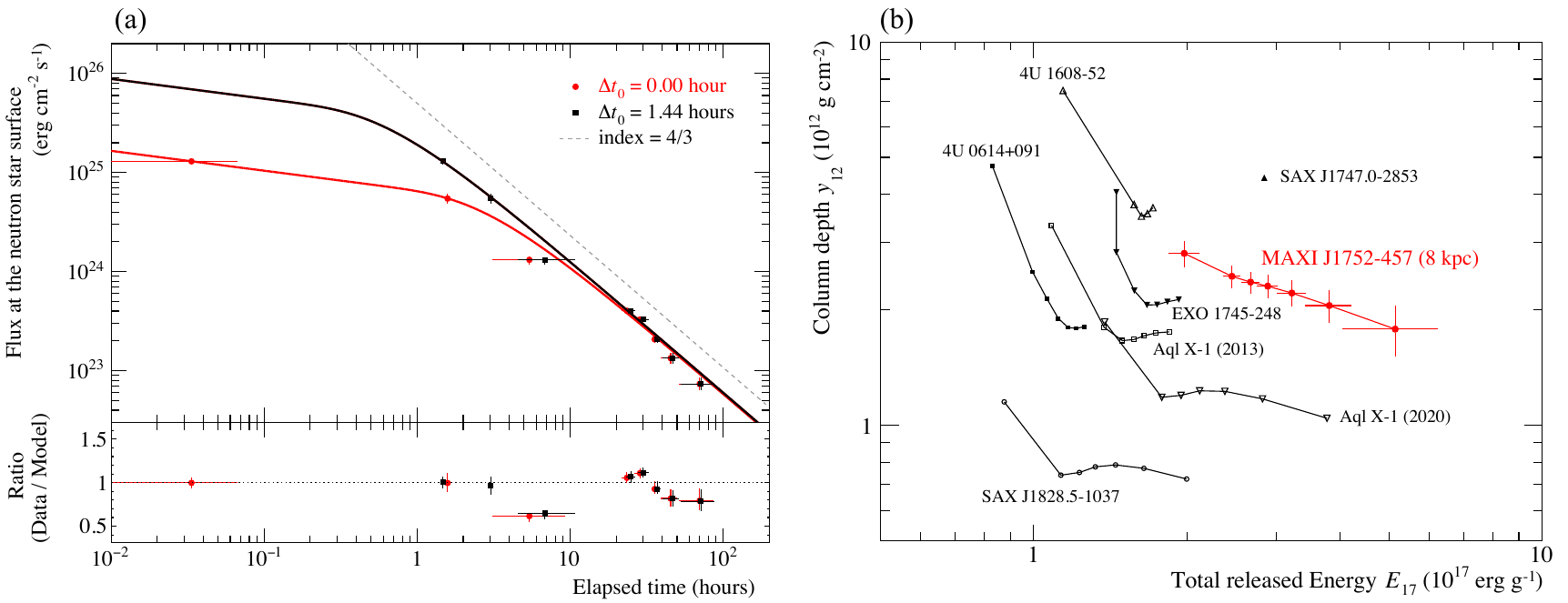}
\caption{
(a) Evolution of the bolometric blackbody flux at the NS surface assuming the source distance of \target at 8~kpc in the log-log format. 
The blackbody parameters are obtained by the 2--10~keV fitting (Figure~\ref{fig:spec} and Table~\ref{tab:spectral_fitting}). 
The observed data are shown in red and black marks, corresponding to the assumptions of the burst onset time at $\Delta t_{0}= 0.00$~hour (i.e., the first MAXI scan of the detection) and $\Delta t_{0}=1.44$~hours, respectively. 
The analytical models are overlaid at the best-fit parameters in Eq. \ref{eq:cm04} of $E_{\textrm{17}}=2.0$ and $y_{12}=7.5$ for the $\Delta t_0=0.00~{\rm hour}$ (red marks), whereas $E_{\textrm{17}}=5.6$ and $y_{12}=5.1$ for the $\Delta t_0=1.44~{\rm hours}$ (black marks). Each reduced $\chi^2$ is 3.16 (d.o.f.$=$6) and 3.50 (d.o.f.$=$6), respectively.
(b) Comparison of the observed burst energy $E_{17}$ and column depth $y_{12}$ of \target (red symbol) with previously reported long-duration bursts (black symbols) \citep{Serino2016}.
In the same way as \citet{Serino2016}, each data point of \target corresponds to the different assumptions of the burst onset at $\Delta t_0=$ 0.00, 0.24, 0.48, 0.72, 0.96, 1.20, and 1.44~hours. 
\label{fig:discussion}}
\end{figure}

In Figure~\ref{fig:discussion}a, we fitted the observed bolometric blackbody flux at the NS surface ($d=8$~kpc) by the analytical cooling model in Eq. \ref{eq:cm04}.
Here, we fixed the slopes at the standard value of $\alpha_{\textrm{f}}=0.2$ and $\alpha_{\textrm{s}}=4/3$. When assuming the burst onset $t_0=T_0$, the decay trend was explained by the two-decay components with the best-fit parameters of $E_{17}=2.0\pm 0.1$ and $t_{\textrm{cool}}=7.7\pm 1.1~{\textrm{hours}}$, the latter of which corresponds to $y_{12}=2.8\pm0.2$. However, since we do not know the true burst onset as discussed in \S\ref{sec:Analysis}, we further scan the burst onset time $t_0=T_0 -\Delta t_0$, ranging in $\Delta t_0=0.00$--$1.44~{\rm hours}$, similar to the discussion in \citet{Serino2016}. 
For example, at $\Delta t_0=1.44~{\textrm{hours}}$, only a single decay feature, proportional to $t^{-4/3}$, is reproduced, suggesting that heat diffusion inside the NS had already finished before our observations (i.e., $\Delta t_0\gtrsim t_{\textrm{cool}}$).

Figure~\ref{fig:discussion}b shows the $E_{17}$--$y_{12}$ relations obtained from the light curve fitting of Figure~\ref{fig:discussion}a with the use of Eq.~\ref{eq:cm04} assuming the source distance at 8~kpc. 
In this plot, we scanned the true burst onset time with $\Delta t_0$ ranging in $0.00$--$1.44$~hours. 
The derived range of $E_{17}$ and $y_{12}$ is consistent with those of other long X-ray bursts previously reported by \citet{Serino2016}. 
The obtained $E_{17}=2.0$--$5.1$ is consistent with the carbon-burning scenario in theory, i.e., $E_{17}=2.1$ for carbon burning~\citep{2003ApJ...583L..87S}, but inconsistent with $E_{17}=15.1$ for pure He burning assuming final products of ${}^{56}{\rm Ni}$. 
The column density at the carbon ignition is estimated at $y_{12}=1.6$--$2.7$, corresponding to $t_{\rm cool}=5.5$--7.7~hours. The total burst energy, estimated from $E_{17}$ and $y_{12}$ at $d=8~{\rm kpc}$, is $E_{\rm tot}=\left(5.3\pm0.7\right)\times10^{42}~{\rm erg}$ for $\Delta t_0=0.00~{\textrm{hours}}$ and $E_{\rm tot}=\left(8.8\pm2.7\right)\times10^{42}~{\rm erg}$ for $\Delta t_0=1.44~{\textrm{hours}}$. 
Our $y_{12}$ and $E_{\textrm{tot}}$ values are qualitatively consistent with superbursters observed so far~\citep{2023MNRAS.521.3608A}. Currently, only two sources, EXO 1745$-$248~\citep{2012MNRAS.426..927A} and SAX J1828.5$-$1037~\citep{Serino2016}, have observed a superburst occurring at very low accretion rates. The mechanism of such superbursts, i.e., how carbon is ignited in cold accreting NSs, remains poorly understood and is an important topic, but this is beyond the scope of this letter, and we leave it to future study.

When making the slopes ($\alpha_{\rm f}$ and $\alpha_{\rm s}$) free in Eq. \ref{eq:cm04_b}, we derived parameters of $\alpha_{\rm f}=0.9 \pm 0.2$ and $\alpha_{\rm s}=1.7 \pm 0.7$ at $\Delta t_0=0.00~{\rm hour}$, corresponding to $p_{\rm f}=0.88\pm0.04$ and $p_{\rm s}=3.1\pm0.3$ in our previous fitting in \S\ref{sec:Analysis}. 
While $\alpha_{\rm s}=1.7 \pm 0.7$ value agrees with the standard cooling model~\citep{Cumming2004}, $\alpha_{\rm f}=0.9\pm 0.2$ value is steeper than the ordinal value $\alpha_{\rm f} =0.2$. This may be due to the significant contribution of temperature-dependent opacity, which mainly consists of electron conductive opacity $\kappa_{\rm cond}$, radiative opacities of electron scattering $\kappa_{\rm es}$, and free-free absorption $\kappa_{\rm ff}$~\citep{1999ApJ...524.1014S}. Assuming $t\lesssim t_{\rm cool}$ in Eq. \ref{eq:cm04} and considering total opacity $\kappa$, the flux is proportional to $E_{17}^{5/3}\kappa^{-2/3}\left(t-t_0\right)^{-1/3}$~\citep{2006ApJ...646..429C}\footnote{Although this formula is derived under assumption of constant opacity, we expect that it holds for the qualitative discussion under small deriation from constant opacity due to temperature dependence (see also Appendix A in \citet{2006ApJ...646..429C}).}. To account for the large $\alpha_{\rm f}$, which implies shorter $t_{\rm cool}$ (i.e., smaller $\kappa$), $\kappa$ should increase with time. This means that $\kappa$ should decrease with temperature. Since $\kappa_{\rm cond}\propto T^2$, $\kappa_{\rm es}\sim \textrm{const}(T)$, and $\kappa_{\rm ff}\propto T^{-7/2}$, radiation of free-free absorption must play a significant role in reproducing the large $\alpha_{\rm f}$ value in MAXI J1752$-$457. For a more precise description of decay indices, NS microphysics such as heating and neutrino cooling processes must be important, although the connection between the microphysics and behavior of fast-decaying light curves is poorly known (e.g., \citealt{2021MNRAS.500.4491Y}). Further numerical studies to construct thermal evolution models with large $\alpha_{\textrm{f}}$ are left for our future work.

\clearpage

\begin{acknowledgments}
We thank T. Kawamuro and M. Shidatsu for their helpful comments on TDE and LMXBs. This project was supported by JSPS KAKENHI (JP23KJ1964, JP17K18776, JP18H04584, JP20H04743, JP20H05648, JP21H01087, JP23K19056, JP24H00008, JP24K00673). T.E. was supported by the “Extreme Natural Phenomena” RIKEN Hakubi project, and the JST Japan grant number JPMJFR202O (Sohatsu).

\end{acknowledgments}

\vspace{5mm}
\facilities{NinjaSat, MAXI}

\bibliography{reference}{}
\bibliographystyle{aasjournal}

% ===========

\newpage

\newpage
\begin{table}[h!]
\centering
\footnotesize
\caption{Summary of the spectral fitting of \target observed with MAXI and NinjaSat.}
\label{tab:spectral_fitting}
\begin{tabular}{ccc|cccc|ccc|cc}
\hline %\parbox[t]{3cm}{Flux \\ ($\times 10^{-9}$ ergs~s$^{-2}$~cm$^{-2}$)} 
\multirow{3}{*}{\parbox[c]{0.4cm}{\centering Time span}} & \multirow{3}{*}{\parbox[c]{2.4cm}{\centering MJD}} & \multirow{3}{*}{\parbox[c]{0.6cm}{\centering Obs.}} & 
\multicolumn{4}{c|}{\tt bbody} &
\multicolumn{3}{c|}{\tt diskbb} & 
\multicolumn{2}{c}{\tt powerlaw} \\
 &  &  & 
\multirow{2}{*}{\parbox[c]{1cm}{\centering Flux}} & \multirow{2}{*}{\parbox[c]{1cm}{\centering $kT$ \\ (keV)}} & \multirow{2}{*}{\parbox[c]{1cm}{\centering $R$ \\ (km)}} & \multirow{2}{*}{\parbox[c]{1.1cm}{\centering $\chi^2$/d.o.f.}}   & 
\multirow{2}{*}{\parbox[c]{1cm}{\centering $kT_{\textrm{in}}$ \\ (keV)}} & \multirow{2}{*}{\parbox[c]{1cm}{\centering $R_{\textrm{in}}\cos{\theta}$\\ (km)}} & \multirow{2}{*}{\parbox[c]{1.1cm}{\centering $\chi^2$/d.o.f.}} & 
\multirow{2}{*}{\parbox[c]{1cm}{\centering $\Gamma$}} & \multirow{2}{*}{\parbox[c]{1.1cm}{\centering $\chi^2$/d.o.f.}} \\
 & & & & & & & & &  \\
\hline 
 A & 60623.77 & M & 21$^{+1}_{-1}$ & $1.78^{+0.10}_{-0.09}$ & $11.2^{+1.2}_{-1.0}$ & $14.3/26$ & $3.4^{+0.3}_{-0.3}$ & $7.7^{+1.5}_{-1.1}$ & $17.7/26$ & $1.6^{+0.1}_{-0.1}$ & $36.2/26$ \\
 B & 60623.83 & M & 9.0$^{+1.0}_{-1.0}$ & $1.7^{+0.2}_{-0.2}$ & $8.2^{+1.6}_{-1.2}$  & $4.3/11$ & $3.1^{+0.5}_{-0.4}$ & $6.1^{+2.1}_{-1.5}$ & $2.8/11$ & $1.7^{+0.2}_{-0.2}$ & $8.6/11$ \\
 C & 60623.85--60624.40 & M & 2.2$^{+0.2}_{-0.3}$ & $1.3^{+0.1}_{-0.1}$ & $7.2^{+1.6}_{-1.2}$  & $7.0/8$ & $2.2^{+0.5}_{-0.4}$ & $6.1^{+2.6}_{-1.9}$ & $6.9/8$ & $1.8^{+0.3}_{-0.3}$ & $8.2/8$ \\
 D & 60624.40--60624.85 & N & 0.66$^{+0.04}_{-0.04}$ & $1.12^{+0.07}_{-0.07}$ & $5.0^{+0.7}_{-0.6}$  & $18.9/18$ & $1.8^{+0.2}_{-0.2}$ & $5.0^{+1.1}_{-0.9}$ & $18.1/18$ & $2.2^{+0.2}_{-0.1}$ & $26.6/18$ \\
 E & 60624.85--60625.10 & N & 0.54$^{+0.03}_{-0.03}$ & $0.99^{+0.05}_{-0.05}$ & $5.7^{+0.7}_{-0.6}$  & $14.2/18$ & $1.5^{+0.1}_{-0.1}$ & $6.5^{+1.1}_{-0.9}$ & $11.6/18$ & $2.5^{+0.1}_{-0.1}$ & $18.6/18$ \\
 F & 60625.10--60625.35 & N & 0.34$^{+0.03}_{-0.03}$ & $0.83^{+0.07}_{-0.06}$ & $6.5^{+1.4}_{-1.1}$  & $10.2/12$ & $1.2^{+0.2}_{-0.1}$ & $8.9^{+2.7}_{-2.2}$ & $12.7/12$ & $2.8^{+0.3}_{-0.3}$ & $20.1/12$ \\
 G & 60625.35--60626.90 & N & 0.22$^{+0.03}_{-0.03}$ & $0.80^{+0.09}_{-0.08}$ & $5.6^{+1.5}_{-1.2}$  & $11.1/12$ & $1.1^{+0.2}_{-0.2}$ & $7.5^{+3.5}_{-2.3}$ & $11.3/12$ & $2.9^{+0.3}_{-0.3}$ & $13.9/12$ \\
 H & 60625.90--60628.40 & N & 0.12$^{+0.02}_{-0.02}$ & $0.60^{+0.06}_{-0.06}$ & $7.3^{+2.1}_{-1.5}$  & $1.8/3$ & $0.8^{+0.1}_{-0.1}$ & $11.9^{+6.0}_{-3.4}$ & $1.8/3$ & $3.7^{+0.4}_{-0.4}$ & $3.5/3$ \\ 
\hline 
\multicolumn{12}{p{50em}}{NOTE. MJD is shown as the elapsed time since the onset of the brightening at MJD 60623.77. ``Obs." means the observatory used for the spectral analysis, where ``M" and ``N" correspond to MAXI and NinjaSat, respectively. Bolometric flux (10$^{-9}$ erg~s$^{-1}$~cm$^{-2}$) in the 0.1--100~keV band is the values at the blackbody fitting. The source distance at 8~kpc is assumed to derive the radii for the blackbody or diskbb models. Each A and B observation is centered at this time with the duration of 50~s.}

\end{tabular}
\end{table}
\end{document}